\documentclass[aps,pra,twocolumn,floats,showpacs]{revtex4}
\usepackage{graphicx}
\usepackage{amsmath}
\usepackage{amssymb}
\usepackage[usenames,dvipsnames]{color}
\usepackage{textcomp}
\usepackage[normalem]{ulem}
\usepackage{bm}
\usepackage{xfrac}

\renewcommand{\a}{{\alpha}}

\newcommand{\E}{{\cal E}}

\newcommand{\D}{{\Delta}}

\newcommand{\T}{{\cal T}}
\newcommand{\Trev}{{T_\text{rev}}}
\newcommand{\Jloc}{{J_\text{loc}}}
\newcommand{\Jdiff}{{J_\text{diff}}}
\newcommand{\Jc}{{J_\text{c}}}

\newcommand {\ket}[1]{|\,{#1}\,\rangle}

\newcommand{\beq}{\begin{equation}}
\newcommand{\eeq}{\end{equation}}
\newcommand{\bea}{\begin{eqnarray}}
\newcommand{\eea}{\end{eqnarray}}

\begin{document}
\title{Control of quantum localization and classical diffusion in laser-kicked  molecular rotors}

\date{\today}
\author{M.~Bitter and V.~Milner}
\affiliation{Department of  Physics \& Astronomy and The
Laboratory for Advanced Spectroscopy and Imaging Research
(LASIR), The University of British Columbia, V6T 1Z1 Vancouver, Canada \\}


\begin{abstract}
{
We experimentally study a system of quantum kicked rotors - an ensemble of diatomic molecules exposed to a periodic sequence of ultrashort laser pulses. In the regime, where the underlying classical dynamics is chaotic, we investigate the quantum phenomenon of dynamical localization by means of state-resolved coherent Raman spectroscopy. We examine the dependence of the exponentially localized angular momentum distribution and of the total rotational energy on the time period between the pulses and their amplitude. The former parameter is shown to provide control over the localization center, whereas the latter one controls the localization length. Similar control of the center and width of a nonlocalized rotational distribution is demonstrated in the limit of classical diffusion, established by adding noise to the periodic pulse sequence.}
\end{abstract}

\pacs{05.45.Mt, 05.45.Gg, 33.80.-b, 42.50.Hz}

\maketitle

\section{Introduction}


The periodically kicked rotor is one of the simplest systems whose classical motion exhibits chaotic dynamics, leading to an unbounded diffusive growth of its energy with the number of kicks. In contrast, the energy growth of a quantum kicked rotor (QKR) is determined  by the interference of quantum interaction pathways \cite{Casati1979}. In the quantum limit, the rotational excitation is either enhanced due to quantum resonances \cite{Izrailev1980} or suppressed due to the effect of dynamical localization \cite{Fishman1982, Bluemel1986}, which has been linked to Anderson localization in solids \cite{Anderson1958}. Similarly to the localization of a quantum particle in a disordered one-dimensional lattice, the QKR localizes in a ``rotational lattice'' of angular momentum states.


Experimental work has mostly been conducted in substitute systems, imitating the QKR behavior with Rydberg atoms in microwave fields \cite{Galvez1988, Bayfield1989, Bluemel1991,Frey1999} or ultracold atoms in optical  lattices \cite{Moore1995, Ammann1998, dArcy2001, Sadgrove2005, Ryu2006, Chabe2008}. In a series of recent articles \cite{Floss2012, Floss2013, Floss2014, Floss2015b, Floss2016}, Averbukh and coworkers proposed a strategy to study a number of QKR phenomena in an ensemble of diatomic molecules exposed to a periodic sequence of ultrashort laser pulses. In this system of true quantum rotors, the effects of a quantum resonance \cite{Cryan2009, Zhdanovich2012, Floss2015a}  and  dynamical localization (DL) \cite{Kamalov2015, Bitter2016c, Bitter2016d} have recently been demonstrated. The recovery of classical diffusion under the influence of noise and decoherence has also been shown experimentally with atoms \cite{Bluemel1991, Klappauf1998, Ammann1998, Milner2000, Oskay2003} and molecules \cite{Bitter2016c}.


The discreteness of the QKR spectrum results in periodic dynamics with a revival time $\Trev=(2cB)^{-1}$, determined by the rotational constant $B$ of the molecule, with $c$ being the speed of light. Matching the period $T$ of a pulse sequence with the so-called quantum resonance at $T=\Trev$ enables an efficient excitation of multiple rotational states with growing (from kick to kick) rotational energy.
On the other hand, away from all full and fractional quantum resonances ($T/\Trev \neq p/q$, where $p$ and $q$ are integers) dynamical localization suppresses the rotational energy growth. This quantum suppression of classical diffusion is accompanied by the exponential line shape of the kicked rotor's quasienergy eigenstates in the angular momentum space. The shape of the observable wave packet depends on the overlap between these quasienergy states, whose spectrum is dictated by the pulse train period, and the initial rotational distribution of molecules \cite{Floss2013}.
In a recent experimental demonstration of DL in laser-kicked molecules, we showed an exponentially localized wave function near the (initially most populated) rotational ground state \cite{Bitter2016c}. We also demonstrated a shifting of the localization center towards higher energies by applying a periodic pulse train to a previously prepared coherent superposition of rotational states and controlling the relative phases between the states \cite{Bitter2016d}.


Here, we study an alternative route to control the final rotational distribution by varying the period of the applied sequence of laser pulses. Oxygen molecules, cooled down to 25~K in a supersonic expansion, are exposed to a series of 13~laser pulses. We  measure the shape of the created rotational wave packet by means of state-resolved coherent Raman spectroscopy. An exponentially localized distribution of the molecular angular momentum is observed in the case of periodic excitation, while a Gaussian line shape, characteristic of classical diffusion, is found for periodic pulse sequences subject to timing noise. In both cases, we investigate the effect of the pulse train period on the rotational distribution - its center, width and shape.
We demonstrate a controlled shift of the quantum localization center and, similarly, the center of the classical distribution, which depends on the time separation between the pulses in a sequence and the nearest fractional quantum resonances.
The control mechanism is associated with the finite duration of the laser pulses, which results in a partially resonant excitation of molecules away from their ground rotational state.


\section{Theoretical background}

The interaction of a diatomic molecule with a periodic train of $N$ linearly polarized laser pulses, not resonant with any electronic transition, is described by the following Hamiltonian:
\begin{equation}\label{Eq-Hamiltonian}
    \hat{H}=\frac{\hat{J}^2}{2I} - \hbar P \cos^2(\theta) \sum_{n=0}^{N-1} \delta(t-nT),
\end{equation}
where $\theta$ is the angle between the molecular axis and the vector of laser polarization, $\hat{J}$ is the angular momentum operator, $I=\hbar(4\pi cB)^{-1}$ is the molecular moment of inertia and $\hbar$ is the reduced Planck constant. The degree of stochasticity in the dynamics of a kicked rotor is governed by a single parameter $K=\tau P$, where $\tau=\hbar T/I$ is the effective Planck constant and $P= \D\alpha /(4\hbar) \int \E^2(t) dt$ is a kick strength, $\D\alpha$ is the molecular polarizability anisotropy and $\E(t)$ is the temporal envelope of the pulse. All experiments discussed in this report satisfy the condition of $K>5$, for which the phase space of the underlying classical motion is fully chaotic and supports an unbounded diffusive growth of rotational energy \cite{Izrailev1990}.

For the periodically driven quantum system described by Hamiltonian (\ref{Eq-Hamiltonian}), the solutions of the time-dependent Schr\"odinger equation are the well-known Floquet states $\chi_\alpha$, which are periodic in time, $\chi_\alpha(t+T) = \exp(-i\E_\a T/\hbar)\ \chi_\alpha(t)$, up to a phase factor determined by the quasienergy $\E_\a$.
Finding quasienergy states of the quantum kicked rotor is equivalent to solving a one-dimensional tight-binding model of a solid-state lattice, whose sites correspond to the rotational states $J$ \cite{Fishman1982}. Owing to the selection rules for a rotational Raman process, laser kicks couple only states of the same parity, resulting in an effective ``rotational lattice'' with a lattice constant  $|\Delta J|=2$. The on-site energy is \cite{Floss2013}
\begin{equation}
    \T_J=\tan( \phi_J)=\tan\left( \frac{\E_\a-E_J}{2\hbar}T\right) \ ,
\label{Eq-TJ}
\end{equation}
where $E_J=hcBJ(J+1)$ is the energy of a molecule, which can be approximated by a rigid rotor due to the low degree of rotational excitation considered here. The coupling between different lattice sites is defined by the kick strength $P$, as described in detail in Ref.\citenum{Floss2013}.

The solution of the rotational tight-binding model critically depends on the site-to-site energy variation, which in turn is determined by the period of the pulse train. It is instructive to express $\phi_J$ through the molecular revival time $T/\Trev$:
\begin{equation}
    \phi_J=\frac{\pi}{2}\left( \frac{\E_\a}{hcB}-J(J+1)\right)\frac{T}{\Trev} \ .
\label{Eq-phiJ}
\end{equation}
On quantum resonance ($T=\Trev$), the on-site energy does not depend on the site number, $\T_J=\tan( \frac{\pi}{2} \frac{\E_\a}{hcB}  )$. All quasienergy states in this \textit{periodic} lattice are \textit{extended states}. The rotational population will therefore spread along the lattice, until it is stopped by the centrifugal distortion which destroys the periodicity \cite{Floss2014,Floss2016}. The same phenomenon persists in a weaker form at fractional resonances ($T/\Trev=p/q$) \cite{Floss2015b}, i.e. the on-site energies of the rotational lattice remain periodic with a period proportional to $q$. If, however, all resonances are avoided, the argument of the tangent function in Eq.(\ref{Eq-TJ}) changes from site to site in a way that makes the energy $\T_J$ a \textit{pseudo-random} function of $J$. The quasienergy states of such a disordered lattice are \textit{localized states} \cite{Fishman1982}.

The molecular kicked rotor is unique in the way the disorder strength depends on the rotational quantum number $J$. While Anderson localization in real space is typically studied in a lattice with spatially uniform disorder, the randomness of the rotational lattice depends on the site number. Importantly, this dependence can be controlled by the excitation period $T$ through its proximity to quantum resonances, as we demonstrate in this work.

\section{Experimental details}

\begin{figure}
\centering
 \includegraphics[width=0.9\columnwidth]{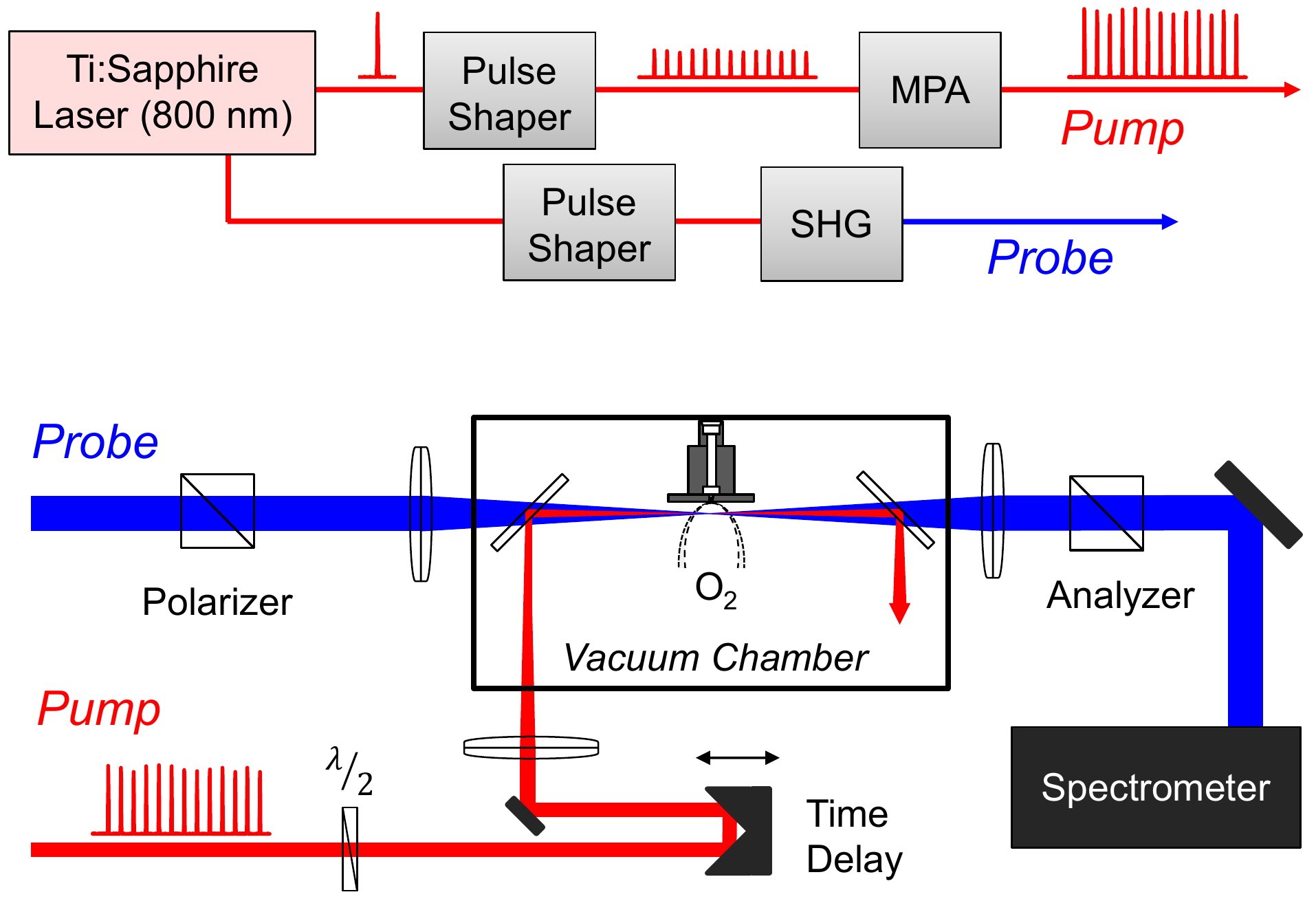}
     \caption{ \textit{(top)} Diagram of the pump and probe sources. Sequences of 13~pulses are generated by a pulse shaper; their energy is then boosted by a multi-pass amplifier (MPA). Another pulse shaper is used to narrow the spectral bandwidth of the probe pulse, whose central wavelength is shifted by means of second harmonic generation (SHG) in a nonlinear crystal.
     \textit{(bottom)} Scheme of the experimental setup. A train of strong femtosecond pulses (pump) and a delayed weak probe pulse are focused on a supersonic jet of oxygen molecules in a vacuum chamber. The change of probe polarization is analyzed as a function of wavelength by means of two crossed polarizers and a spectrometer. }
  \vskip -.1truein
  \label{Fig:Setup}
\end{figure}

A high-intensity train of femtosecond pulses is generated in an optical system \cite{Bitter2016a}, shown schematically at the top of Fig.~\ref{Fig:Setup}.
We use a Ti:Sapphire laser system producing pulses of 130~fs full width at half maximum (FWHM) at a central wavelength of 800~nm, 1~kHz repetition rate, and 2~mJ per pulse. Part of the beam ($40\%$ in energy) is sent through a standard `$4f$' pulse shaper \cite{Weiner2000}, which generates a sequence of 13 pulses of equal amplitudes separated by arbitrary time intervals in a total window of 50~ps. The revival time of oxygen, $T_\text{rev}\approx 11.67$~ps, is long enough for a 130~fs pulse to act as a $\delta$-kick, yet short enough to generate sufficiently many pulses within the maximum time window accessible by the pulse shaper. The pulse sequence is amplified by a home-built multi-pass amplifier to reach a kick strength of up to $P=8$ per pulse ($\sim 3\cdot 10^{13}~\mathrm{W/cm^2}$) at 10~Hz repetition rate. The standard deviation of the pulse energy fluctuations is about $15\%$. The remaining part of the 800~nm beam is used as a probe. Its spectrum is narrowed down in a separate pulse shaper, before its central wavelength is shifted to $\approx 400$~nm by means of second harmonic generation in a nonlinear optical crystal.

The experimental setup is illustrated at the bottom of Fig.~\ref{Fig:Setup}. The probe pulse of 0.15~nm spectral width (FWHM) is linearly polarized at 45\textdegree\  with respect to the pulses in the pump train. Both beams are focused into a vacuum chamber, where they are combined on a dichroic beamsplitter and intersect a supersonic jet of oxygen molecules. The pump pulses produce coherent molecular rotation, which modulates the refractive index of the gas. As a result, Raman sidebands appear in the narrow-band spectrum of the weak probe pulse, polarized orthogonally to its initial polarization \cite{Korech2013, Korobenko2014a}. Each Raman peak is shifted from the central probe frequency by an amount that depends on the rotational quantum number $J$, while its intensity $I_J$ allows the retrieval of the corresponding population $P_J$ \cite{Bitter2016c}. Critical to these experiments, we achieve a dynamic range of four orders of magnitude in Raman intensity (two orders in rotational population).
Special care is taken to avoid detrimental effects of spatial averaging by making the probe beam significantly smaller than the pump (FWHM beam diameters of $20~\mathrm{\mu m}$ and $60~\mathrm{\mu m}$, respectively). We use a $500~\mathrm{\mu m}$ diameter pulsed nozzle, operating at the repetition rate of 10~Hz and the stagnation pressure of 33~bar, to lower the rotational temperature of oxygen down to 25~K at a distance of 2~mm from the nozzle.


\section{Results}
\subsection{Dynamical localization}


\begin{figure}
\centering
 \includegraphics[width=0.85\columnwidth]{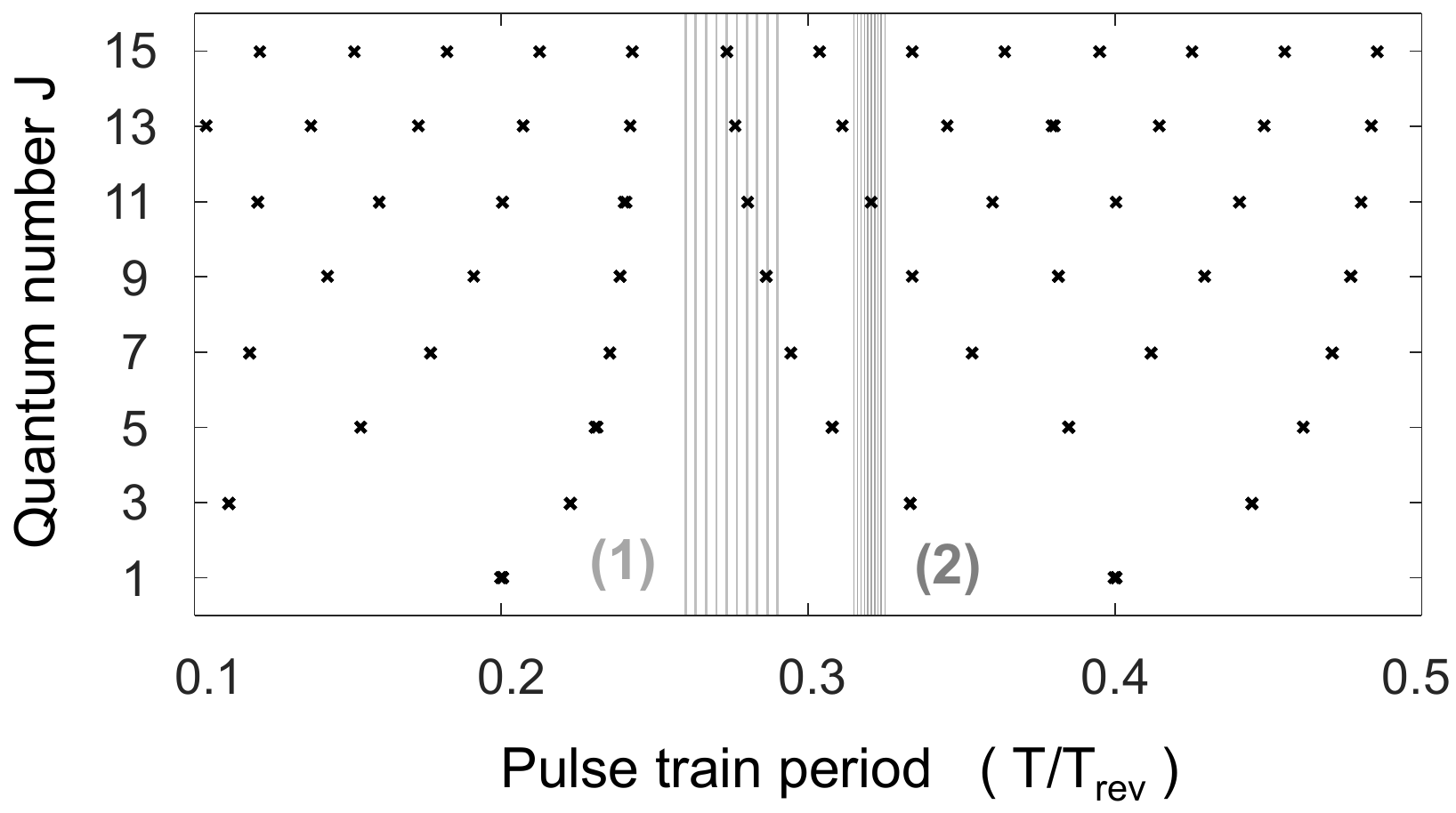}
     \caption{ Rotational resonance map for $^{16}$O$_2$. Markers indicate the values of $T$, for which the relative phase between the two neighboring sites $J$ and $J+2$ of the rotational lattice is equal to an integer multiple of $\pi$. Two sets of ten vertical lines each indicate the experimental pulse train periods, (\textbf{1}) $0.26\leqslant T/\Trev \leqslant 0.29$  and (\textbf{2}) $0.315\leqslant T/\Trev \leqslant 0.325$. }
  \vskip -.1truein
  \label{Fig:Map_periodic}
\end{figure}

Observing the effect of dynamical localization relies on avoiding fractional quantum resonances. Figure~\ref{Fig:Map_periodic} shows the location of resonances (crosses) and illustrates our strategy in choosing the appropriate pulse train periods (vertical lines). Each cross represents the value of $T$, for which the relative phase between the two neighboring sites $J$ and $J+2$ of the rotational lattice, $\Delta\phi_J=\phi_{J+2}-\phi_J=\pi(2J+3)T/\Trev$, is equal to an integer multiple of $\pi$, meaning that the two on-site energies are the same. (Equivalently, these are also the times when a rotational wave packet, consisting of the states $\ket{J}$ and $\ket{J+2}$, completes an integer number of half-rotations). Note, that only odd rotational states are allowed because of the nuclear spin statistics of $^{16}$O$_2$.


\begin{figure}
\centering
 \includegraphics[width=0.9\columnwidth]{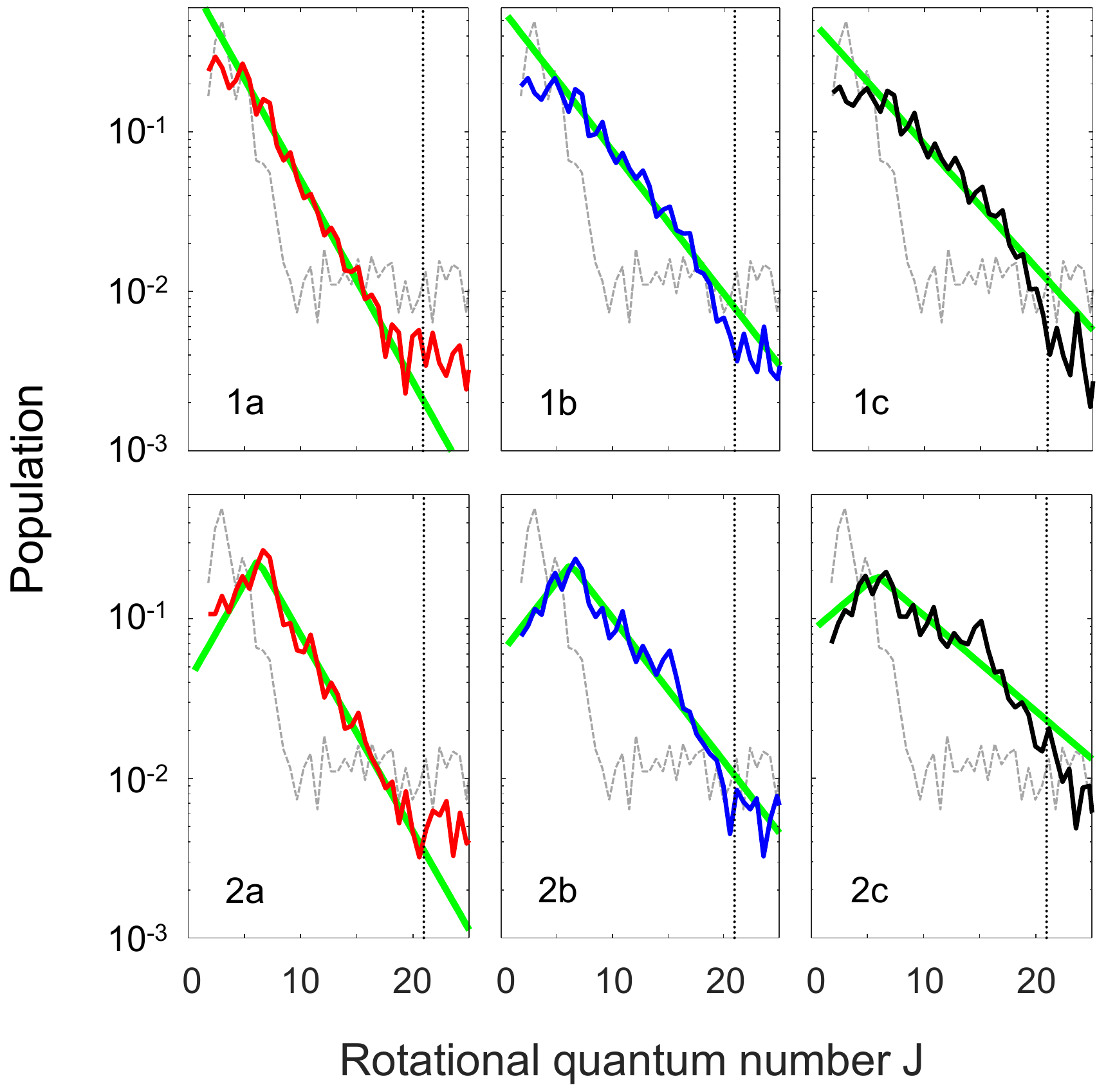}
     \caption{ (color online) Localized angular momentum distributions of oxygen molecules excited with a periodic train of 13~pulses for three kick strengths of (\textbf{a}) $P=4$, (\textbf{b}) $P=6$ and (\textbf{c}) $P=8$.
    Shown are the average distributions obtained from ten pulse trains with the mean period of (\textbf{1}) $\bar{T} =0.275\ \Trev$ and (\textbf{2}) $\bar{T}=0.32\ \Trev$, see Fig.~\ref{Fig:Map_periodic}.
    The populations are fitted with an exponential function (thick green line) and compared to the initial thermal distribution at $T=25$~K (dashed grey line). The dotted vertical line represents the excitation limit due to the finite pulse duration.
}
  \vskip -.1truein
  \label{Fig:Population_periodic}
\end{figure}

We start by choosing a set of ten evenly spaced values of $T$ in the interval $0.26\leqslant T/\Trev \leqslant 0.29$, marked with (\textbf{1}) in Fig.~\ref{Fig:Map_periodic}. Here, all resonances belonging to low-lying $J$-states, which are initially populated, are far off-resonant.
The upper row in Fig.~\ref{Fig:Population_periodic} shows the measured population distribution averaged over these ten \textit{periodic} trains of 13 pulses each. Plots (\textbf{1a-1c}) correspond to the individual kick strength of $P=4,6,8$, respectively. The rotational population has been extracted from the detected Raman signal as $P_J = a \sqrt{I_J}$, with the coefficient $a$ found from normalizing the total population to unity. We note, that this method of calculating $P_J$ neglects the initial distribution of molecules among more than one rotational state, including the degenerate magnetic sub-levels, which results in a systematic underestimation of the amount of molecules at low $J$'s \cite{Bitter2016c}. In the analysis of our results below, we therefore fit the extracted distributions starting from $J\geqslant 4$ and up to the level containing more than 1\% of the total population. The Raman frequency shift (horizontal axis) has been converted to the rotational quantum number $J$.

In Fig.~\ref{Fig:Population_periodic} (\textbf{1a}), the angular momentum distribution decays exponentially away from the initially populated states centered at $J=1$ down to the instrumental noise floor around $P_J\approx 5\cdot 10^{-3}$. It can clearly be distinguished from the initial Boltzmann distribution (grey dashed line), which has been recorded after a single weak pulse. The evident exponential shape (fit indicated by the thick green line), with a localization length ($1/e$ width) $\Jloc=3.4\pm0.3$, is a hallmark of dynamical localization. Note that an unavoidable (due to the finite pulse length) overlap with higher fractional resonances $J\geq 9$ (Fig.2) is not only weak, but also averages out between different pulse trains, and hence does not distort the overall exponential line shape. For stronger kicks of $P=6$ and 8 (\textbf{1b,1c}), the localization length increases to $\Jloc=4.8\pm0.5$ and $5.6\pm0.6$, respectively. The distributions become sub-exponential at higher values of angular momentum due to the finite duration of the laser pulses (130~fs FWHM). Beyond the limit of $J_\text{lim}= 21$ (i.e. to the right of the dotted vertical line), an oxygen molecule rotates by $\gtrsim 90$\textdegree\ during the length of the pulse, which lowers its effective kick strength and suppresses further rotational excitation. Each exponential fit only includes experimental values up to $J_\text{lim}$.\\


To demonstrate the control over the localized distribution and its localization center, we now select a set of ten evenly spaced pulse train periods in the second interval $0.315\leqslant T/\Trev \leqslant 0.325$, marked with (\textbf{2}) in Fig.~\ref{Fig:Map_periodic}. Although the selected periods are again nonresonant, as required for DL, the proximity of \textit{all} pulse trains to the two low-lying fractional resonances associated with $J=3$ at $T/\Trev=1/3$ and $J=5$ at $T/\Trev=4/13$, alters the shape of the localized distribution. In oxygen, the separation between these two resonances is about 300~fs, such that the finite pulse duration results in a partial overlap with both resonances for all chosen periods. Therefore, a population transfer to higher rotational states is facilitated before the QKR dynamically localizes. This mechanism is unique to finite pulse durations and could not be replicated in our numerical simulations with $\delta $-kicks.

In Fig.~\ref{Fig:Population_periodic} (\textbf{2a-2c}), the localization center $\Jc$ is significantly shifted away from the edge, with its position being independent of the kick strength.  An exponential fit of the form $P_J \propto \exp(-|J-\Jc|/\Jloc)$ yields $\Jc=6.2\pm0.5, 6.2\pm0.7, 5.8\pm1.3$ for $P=4,6,8$, respectively (green lines). In comparison to the far off-resonant case (\textbf{1a-1c}), the different localization lengths $\Jloc=3.5\pm0.4, 4.8\pm0.6, 7.2\pm1.3$ remain the same, as they are solely determined by the kick strengths of the pulse train. Only for $P=8$ (\textbf{2c}) we observe a larger localization length than in (\textbf{1c}), caused by a local flattening of the distribution around $J=11-15$. This deviation reflects the proximity of the train period to the next higher lying resonances (see Fig.~\ref{Fig:Map_periodic}). The effect is more pronounced for stronger kicks when such higher rotational states become populated.

\subsection{Classical diffusion}


Dynamical localization relies on quantum coherences. Both timing and amplitude noise were shown to destroy localization and recover classical diffusion of the QKR angular momentum \cite{Bitter2016c}. Here, we investigate how fractional resonances affect the noise-induced  classical diffusion. To make the proper comparison with the localized scenario, we left the effective Planck constant $\tau$, the kick strength $P$ and, accordingly, the stochasticity parameter $K=\tau  P$ the same.

\begin{figure}
\centering
 \includegraphics[width=0.85\columnwidth]{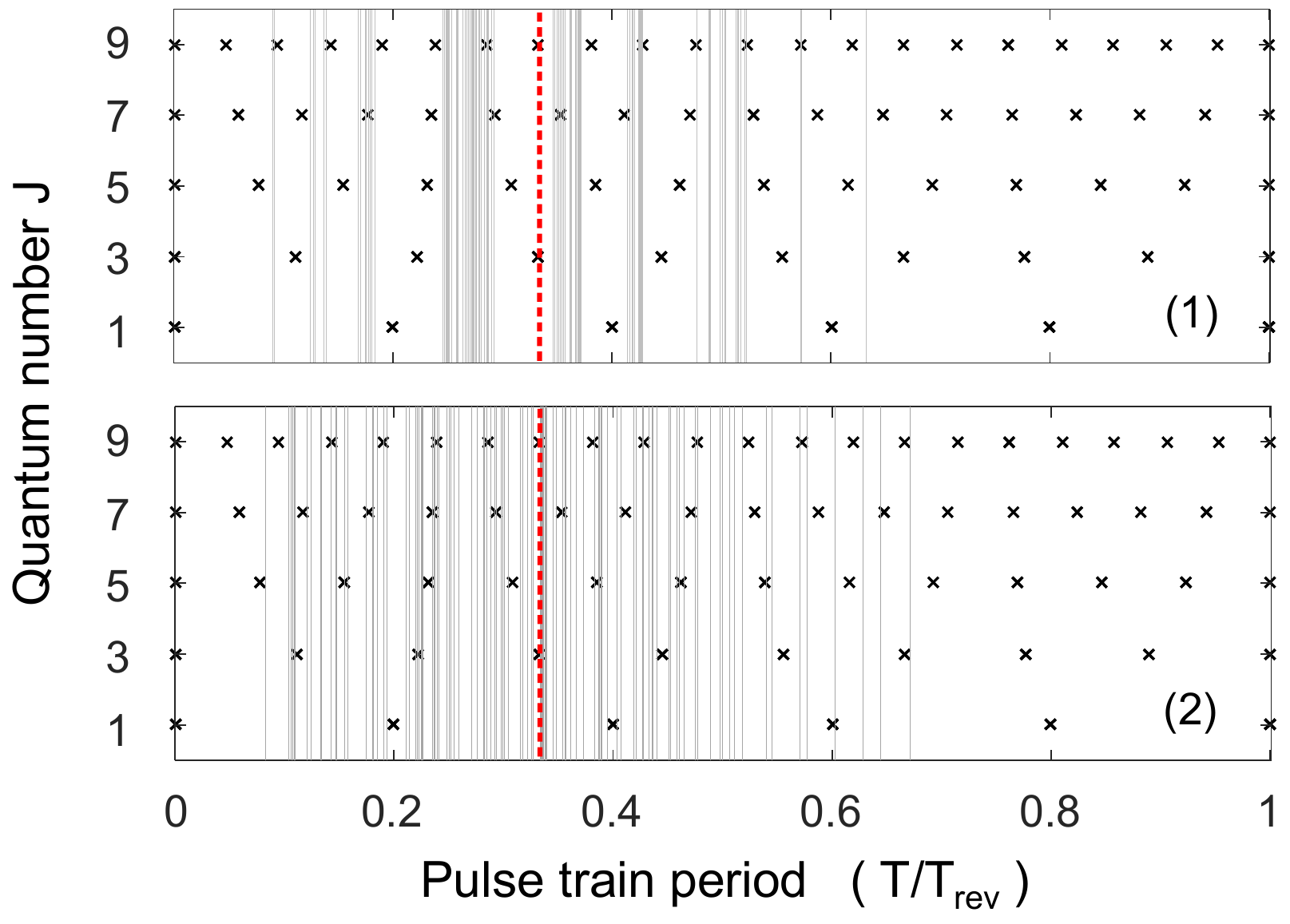}
     \caption{ (color online) Same resonance map as in Fig.\ref{Fig:Map_periodic}.      Here, 120 vertical lines indicate the random periods of ten different pulse sequences, 13~pulses each, following a Gaussian distribution with a mean and a standard deviation of (\textbf{1}) $\bar{T} =0.34\ \Trev$ and $35\%$, (\textbf{2}) $\bar{T} =0.32\ \Trev$ and $43\%$, respectively. In case (\textbf{1}), no period is within 150~fs of any fractional resonance associated with $J=1,3$ or 5. No such restriction was imposed in case (\textbf{2}). The dashed red lines mark the two mean periods.
 }
  \vskip -.1truein
  \label{Fig:Map_random}
\end{figure}

We introduce timing noise by randomly varying the time intervals between the 13~pulses in each train (using the pulse shaping technique, described earlier) with a standard deviation of $\sigma_T$ around the mean period $\bar{T}$, which is chosen to be similar to its value in the case of periodic trains. The upper plot in Fig.~\ref{Fig:Map_random} shows the first set of 120~random periods (for 10 pulse trains), which follow a Gaussian distribution with $\bar{T} =0.34\ \Trev$ and $\sigma_T=35\%$, but are engineered to avoid all quantum resonances with low-lying rotational states $J=1,3$ or 5. In case (\textbf{2}), shown in the lower plot of Fig.~\ref{Fig:Map_random}, the distribution is truly random with no restrictions, $\bar{T} =0.32\ \Trev$ and $\sigma_T=43\%$.

\begin{figure}
\centering
 \includegraphics[width=0.9\columnwidth]{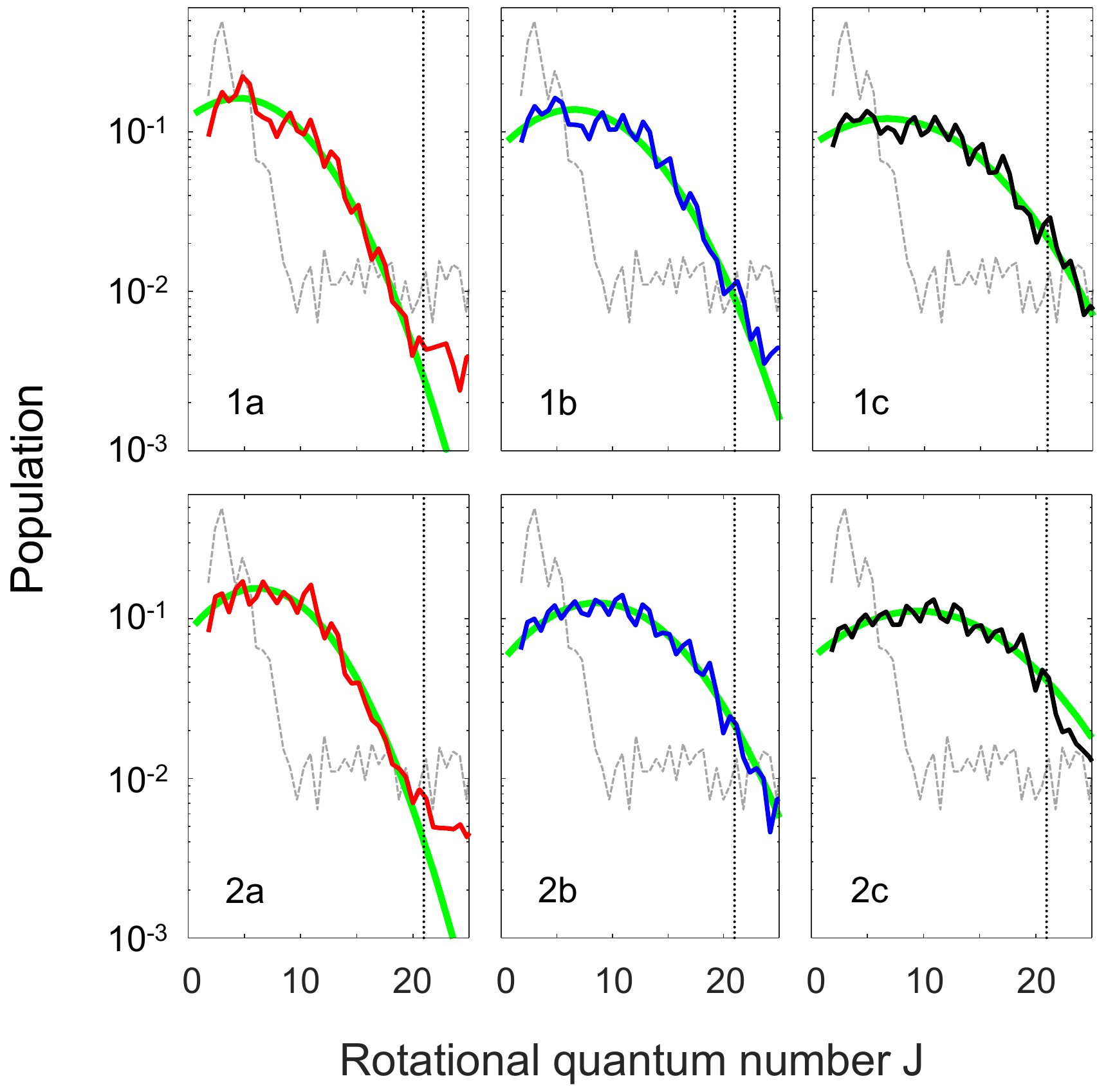}
     \caption{ (color online) Angular momentum distribution of oxygen molecules excited by nonperiodic trains of 13~pulses for three kick strengths of (\textbf{a}) $P=4$, (\textbf{b}) $P=6$ and (\textbf{c}) $P=8$.
    Shown are the average distributions obtained from ten pulse trains with a mean and a standard deviation of (\textbf{1}) $\bar{T} =0.34\ \Trev$ and $35\%$, (\textbf{2}) $\bar{T} =0.32\ \Trev$ and $43\%$, respectively, see Fig.~\ref{Fig:Map_random}. In case (\textbf{1}), no period is within 150~fs of any fractional resonance associated with $J=1,3$ or 5.
    The populations are fitted with a Gaussian function (thick green line) and compared to the initial thermal distribution at $T=25$~K (dashed grey line). The dotted vertical line represents the excitation limit due to the finite pulse duration.
}
  \vskip -.1truein
  \label{Fig:Population_random}
\end{figure}


Figure~\ref{Fig:Population_random} presents the population distributions, obtained from the average over ten \textit{nonperiodic} pulse trains, for the same six cases exhibited in Fig.~\ref{Fig:Population_periodic}. We observe a qualitatively different nonexponential shape, expected for a classical kicked rotor. In contrast to the exponential line shapes of DL, now destroyed by noise, the angular momenta follow a Gaussian distribution (thick green line), which is characteristic of classical diffusion. Its $1/e$ width $\Jdiff$  is wider than the corresponding localized counterpart $\Jloc$ for all six sets of experimental parameters.

In plots (\textbf{1a-1c}), when the random periods are designed to avoid low-lying resonances, a Gaussian fit yields the distribution center at $\Jc=4.4\pm1.6, 6.5\pm1.2, 6.7\pm1.6$ and a width of $\Jdiff=8.2\pm1.1, 8.8\pm1.0, 10.9\pm1.7$ for $P=4,6,8$, respectively. The distribution becomes broader with the increasing kick strength.
In comparison, when the train periods randomly overlap with fractional resonances [plots (\textbf{2a-2c})], we detect a noticeable shift of the Gaussian center towards higher rotational states $\Jc=6.2\pm1.2, 8.7\pm1.2, 9.5\pm1.3$, which becomes larger with the increasing kick strength $P$. The distribution widths of $\Jdiff=7.7\pm0.8, 9.3\pm1.3, 11.4\pm2.2$ remain similar, within the errors of the fit, to the ones measured with set (\textbf{1}) of the noisy pulse trains.

\subsection{Rotational energy}

\begin{figure}
\centering
 \includegraphics[width=0.9\columnwidth]{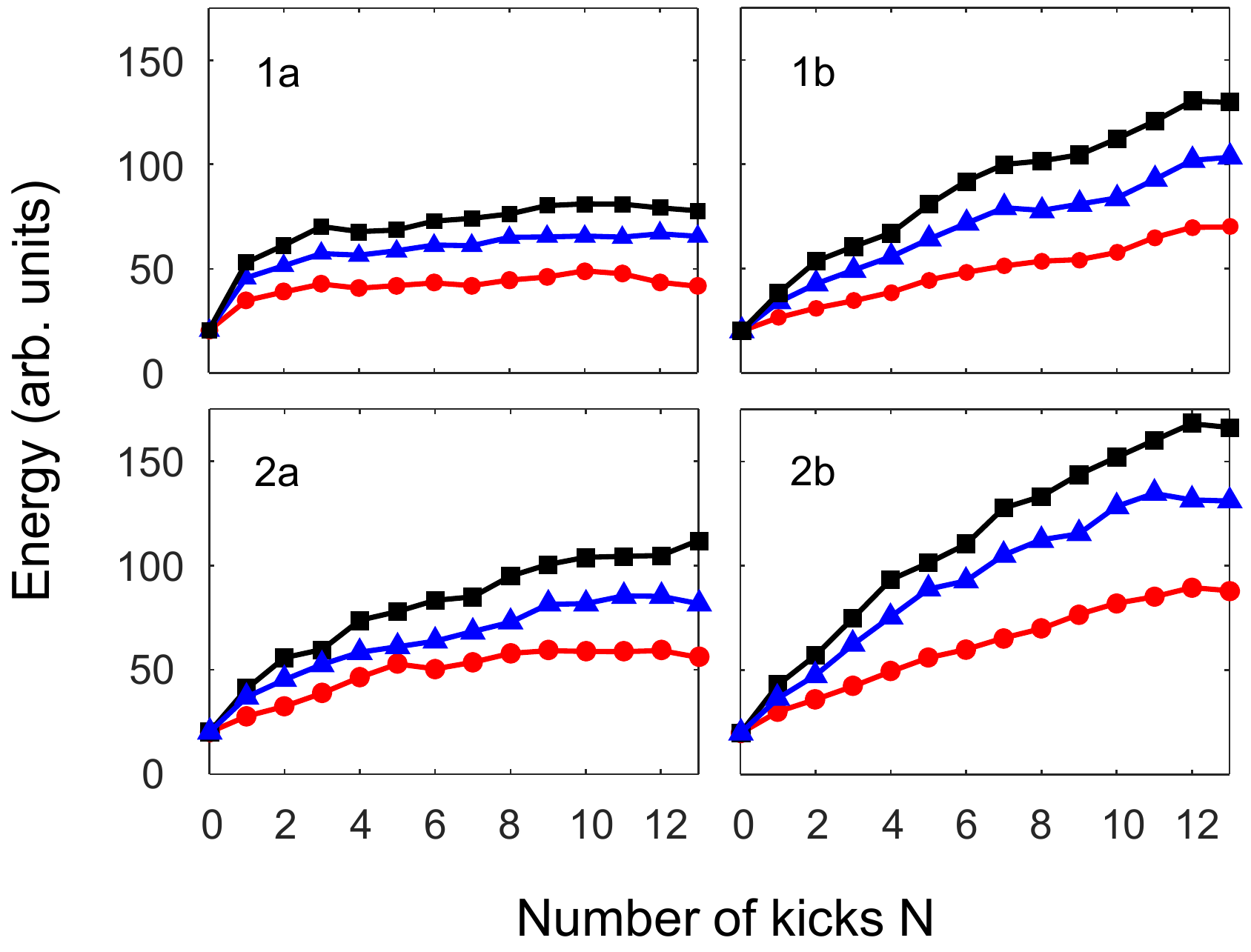}
     \caption{ (color online) Rotational energy as a function of the number of kicks $N$ with a kick strength $P=4$ (red circles), $P=6$ (blue triangles) and $P=8$ (black squares). Compared are the energies for (\textbf{a}) periodic  and (\textbf{b}) nonperiodic sequences, with periods that (\textbf{1}) avoid or (\textbf{2}) allow overlap with low-lying quantum resonances, as shown in Fig.~\ref{Fig:Map_periodic} and Fig.~\ref{Fig:Map_random}. }
  \vskip -.1truein
  \label{Fig:Energy}
\end{figure}

The ability to resolve individual rotational states allows us to determine the rotational energy, absorbed by the molecules, as $\sum_J E_J P_J$. The absorbed energy is plotted in Fig.~\ref{Fig:Energy} as a function of the number of kicks $N$ for all the discussed excitation scenarios. In the case of nonresonant \textit{periodic} pulse trains (\textbf{1a}), the rotational energy ceases to grow after as few as three pulses, after which the dynamical localization sets in. Larger localization lengths for the increasing kick strength, from $P=4$ (red circles) to $P=6$ (blue triangles) to $P=8$ (black squares), are reflected in a higher saturation level of the rotational energy. Periodic pulse trains that promote the initial population transfer via the $J=3$ and 5 resonances lead to a greater absorption of energy (\textbf{2a}). It takes more pulses for the dynamical localization to set in and the total energy to saturate. To observe a completely suppressed energy growth one would need a sequence of more than 13~pulses.

In contrast, \textit{nonperiodic} sequences result in a continuous increase of rotational energy, exceeding the energy achieved by the corresponding periodic sequences. The growth rate is faster for truly random periods (\textbf{2b}) compared to random periods that are tailored to avoid fractional resonances (\textbf{1b}). The continuous absorption of energy and the Gaussian line shape are both manifestations of classical diffusion. The sublinear growth rate is due to the finite duration of the laser pulses, discussed earlier in the text.

\section{Conclusions}

We investigated how the period of a pulse train affects the shape of the angular momentum distribution of a molecular rotor. Off-resonant excitation, which can be mapped onto pseudo-random on-site energies in a rotational lattice, is shown to lead to dynamical localization. Due to the finite duration of the experimental pulses, the shape of the localized distribution is affected by the partial overlap of the train period with fractional quantum resonances. The average distribution obtained from ten periodic sequences with different  nonresonant periods revealed an exponentially localized spectrum with a localization length $\Jloc$ that depended only on the kick strength $P$. We measured $\Jloc=3.4, 4.8, 5.6$ for values of $P=4,6,8$, respectively.
On the other hand, by matching the periodicity of the pulse train to low-lying resonances, we transferred the population to higher rotational states before it dynamically localized. We demonstrated a localized distribution centered around $J\approx6$ with the anticipated localization length according to different kick strengths.

Breaking the periodicity by introducing timing noise destroys the localization of the quantum rotor's wave function and leads to classical diffusion. We showed that the resulting Gaussian population distribution becomes wider with the number of pulses as well as their kick strength. The distribution center is equally affected by the relative position of the random periods with respect to fractional resonances. Omitting resonances resulted in a slower spread to higher $J$-states. The rotational energy of the QKR has been studied as a function of the number of kicks. It reflected both processes, the one of dynamical localization, when the energy growth is shown to cease completely after as few as three pulses, and the one of classical diffusion with an unbounded increase in energy.

The authors would like to thank J.~Flo{\ss} and I.~Sh.~Averbukh for stimulating discussions on the subject of the quantum kicked rotor.


\end{document}